\documentclass[times,11pt]{elsart}

\usepackage{amsmath} 
\usepackage{amssymb} 
\usepackage{amsfonts}
\usepackage{graphicx,color}                 
\usepackage{subfigure}
\usepackage{rotating}
\usepackage{multirow}
\begin{document}

\begin{frontmatter}
\title{Reply to ``Comment on `Ensemble Kalman filter with the unscented transform' ''}

\author[kaust,man]{X. Luo\corauthref{lxd}}
\corauth[lxd]{Corresponding author.}
\ead{xiaodong.luo@kaust.edu.sa}
\author[maths]{, I.M. Moroz}
\author[kaust]{, and I. Hoteit}
\address[kaust]{King Abdullah University of Science and Technology, Thuwal, KSA}
\address[maths]{Mathematical Institute, 24-29 St Giles', Oxford, UK, OX1 3LB}
\address[man]{Oxford-Man Institute of Quantitative Finance, Eagle House, Walton Well Road, Oxford, UK, OX2 6ED}

\begin{abstract}
This is a reply to the comment of Dr Sakov on the work ``Ensemble Kalman filter with the unscented transform'' of Luo and Moroz (2009) \cite{Luo-ensemble}.

\end{abstract}

\end{frontmatter}

\maketitle

\section{Discussion}
Dr Sakov cited an earlier work \cite{Wang-which}, which is similar to \cite{Luo-ensemble}, but whose conclusion contradicts that in \cite{Luo-ensemble}. We note that these conclusions shall be interpreted based on how the filter is implemented. In \cite{Wang-which}, the authors considered the combination of the ensemble transform Kalman filter (ETKF) with the positive-negative pair (PNP) scheme. In order to keep the size of the analysis ensemble constant at each assimilation cycle, the authors in \cite{Wang-which} choose to discard half of the members before propagating them forward (otherwise the ensemble size will double at each cycle). The advantage in doing this is that the computational cost can be very cheap. However, as a potential problem, the sample covariance of the remaining ensemble members might not be a good approximation to the one before halving the size, which may thus deteriorate the performance of the filter.

Dr Sakov also compared the numerical results of \cite{Luo-ensemble} with the existing ones in the literature, for example, \cite{Ott-local,Sakov2008,Whitaker-ensemble}. In terms of absolute rms error, Dr Sakov commented that the results in \cite{Luo-ensemble} are in the range of $0.73-0.99$ \cite{Sakov2009}, while those in \cite{Ott-local,Sakov2008,Whitaker-ensemble} can be as low as $0.17-0.21$. This difference may depend on how the comparison is made. For example, in order to obtain a better performance, in \cite{Ott-local} the authors used a bank of ETKFs, while in \cite{Sakov2008} the ensemble size was up to $91$ for the $40$-dimensional Lorenz-Emanuel model (L40 after \cite{Sakov2009}). In contrast, in \cite{Luo-ensemble} the number of sigma point is at most $13$, since the upper bound $l_u=6$. If in \cite{Sakov2008} one chose the ensemble size to be $13$ for the L40 system, then it can be seen that the rms errors of the various ETKFs are all above \begin{scriptsize}\end{scriptsize}$1$ (cf. Fig. 3 of \cite{Sakov2008}), higher than the range of $0.73-0.99$ achieved by the EnUKF in \cite{Luo-ensemble}.

In \cite{Luo-ensemble}, we have focused on the scenario where the ensemble size (or the rank of the sample covariance) in the filter is typically (much) lower than the dimension of the system in assimilation. Nevertheless, it should be interesting to compare the performances of the EnUKF and the EnKF (with the ETKF as the representative) beyond the aforementioned scenario, where the ensemble size of a filter could be close to, or even higher than, the dimension of the system in assimilation. To this end, in what follows we would like to conduct one more experiment, supplementary to those in \cite{Luo-ensemble}. 

In the experiment, we examine the performances of the EnUKF and the ETKF under various covariance factors (but without covariance filtering to avoid complicating our discussion). The settings in the experiment are as follows. The truths and observations are generated in the same way as that in \cite[\S~4]{Luo-ensemble}. For the EnUKF, we let its intrinsic parameters $\beta=2$, $\lambda=-2$, the threshold $h_1 =1000$. We let the lower bound $l_l$ equal to the upper bound $l_u$ (hence the truncation number $l=l_l=l_u$), and take their values in the set $[5, 10, 15, 20, 30, 40]$, which implies that the corresponding ensemble size $n=2l+1$ takes values in the set $[11, 21, 31, 41, 61, 81]$. In addition, we let the inflation factor $\delta$ take values in the set $[[0.05:0.05:0.5],[1:0.5:6]]$. Here, the notation $[0.05:0.05:0.5]$ denotes the set whose elements increase from $0.05$ to $0.5$, with an even increment of $0.05$ each time. Similar interpretation can also be applied to the notation $[1:0.5:6]$, while $[[0.05:0.05:0.5],[1:0.5:6]]$ means the concatenation of the sets $[0.05:0.05:0.5]$ and $[1:0.5:6]$. For comparison, we use the same ensemble sizes $n$ and inflation factors $\delta$ in the ETKF. At the first assimilation cycle, the EnUKF is always initialized with a $10$-member background ensemble, no matter what the value of $l$ (hence $n$) is. But for the ETKF, it is initialized with an $n$-member background ensemble. To reduce the statistical fluctuation, we repeat the experiment for both the EnUKF and the ETKF for $20$ times, each time with the same initial system state (hence the subsequent truths) and observations, but a randomly drawn initial background ensemble.

\begin{figure*}[!t]
\centering
\includegraphics[width=\textwidth]{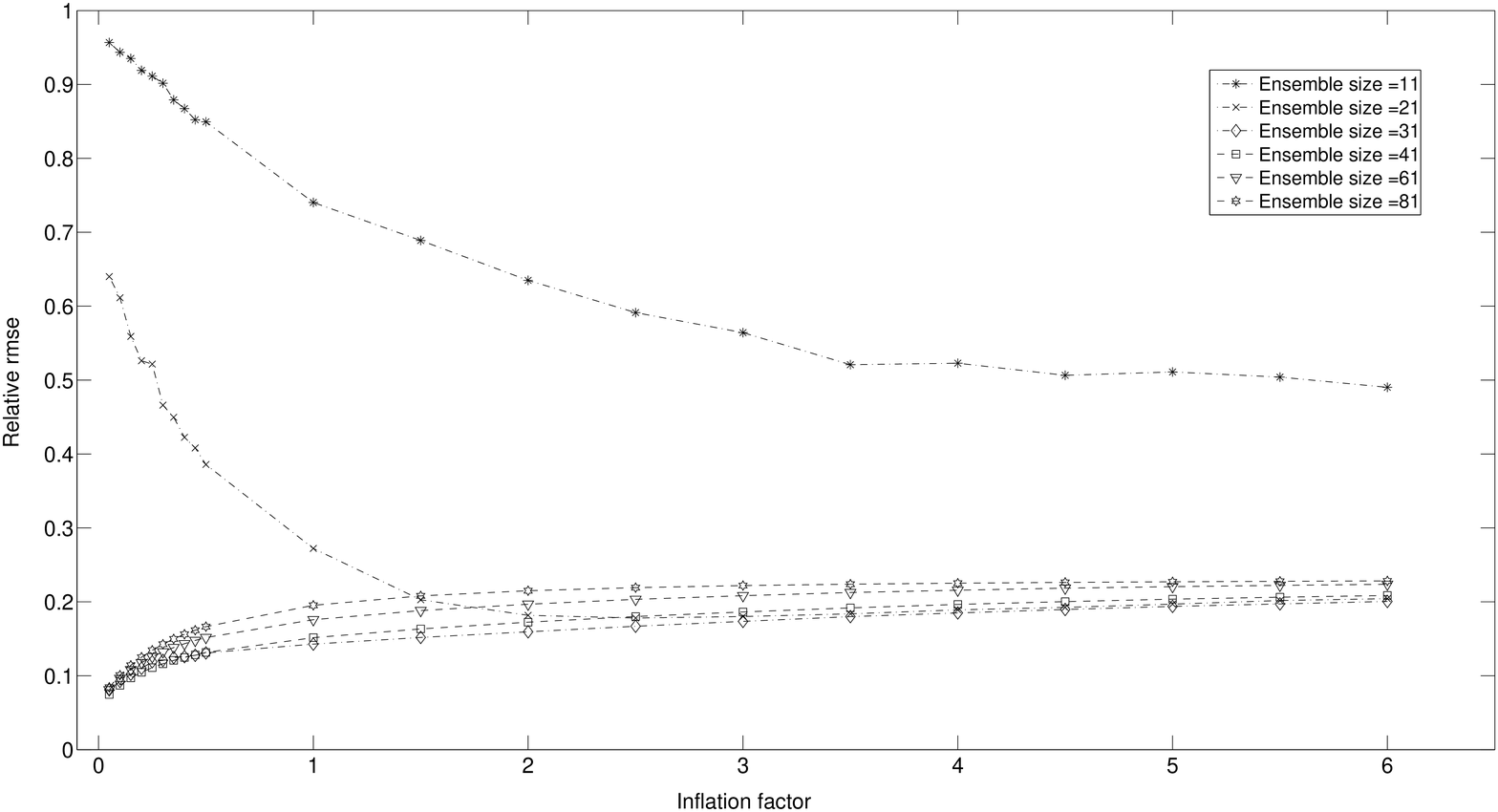}
\caption{ \label{fig:EnUKFL96_noFiltering_varInflationEnSize} Relative rms errors of the EnUKF (with different ensemble sizes) as functions of the inflation factors.}
\end{figure*}  

\begin{figure*}[!t]
\centering
\includegraphics[width=\textwidth]{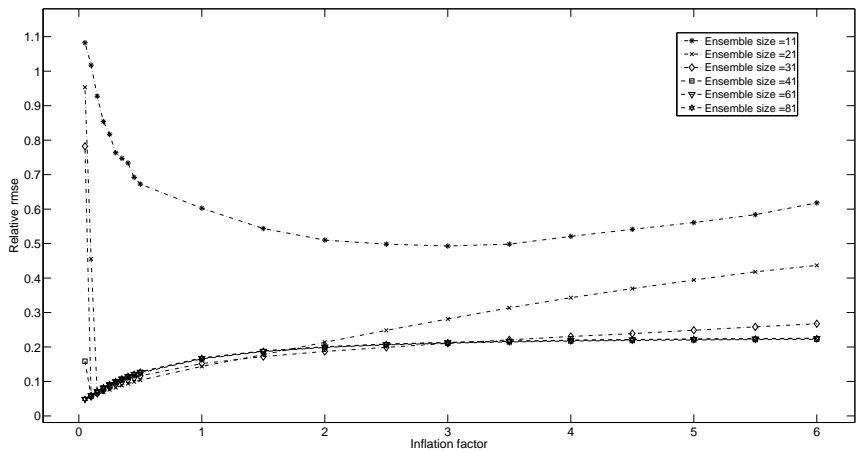}
\caption{ \label{fig:ETKFL96_noFiltering_varInflationEnSize} Relative rms errors of the ETKF (with different ensemble sizes) as functions of the inflation factors.}
\end{figure*}  

In Figs.~\ref{fig:EnUKFL96_noFiltering_varInflationEnSize} and~\ref{fig:ETKFL96_noFiltering_varInflationEnSize}, we show the relative rms errors of the EnUKF and the ETKF (with different ensemble sizes) as functions of the inflation factor. As one can see in them, in order to achieve a relatively good performance, the choice of the inflation factor may be different with different ensemble sizes. In the large sample scenario (say, $n > 40$), a small covariance inflation factor, e.g., $\delta=0.05$, appears good for both the EnUKF and the ETKF, which is consistent with Dr Sakov's comment. However, in the small sample scenario (say, $n=11$), one may need a larger inflation factor, e.g., $\delta=3$ for the ETKF and $\delta=6$ for the EnUKF, to achieve a relatively good result. Our explanation is that in the small sample scenario, there exist relatively large sampling errors. Hence, a larger inflation factor is needed to compensate for the effect of finite ensemble size. 

Given the same ensemble size $n$, in terms of the lowest relative rms errors that the filters can achieve within the tested parameter ranges, the performances of the EnUKF and the ETKF are comparable when $n=11$, with the lowest rmse of the EnUKF being $0.490$, slightly lower than that of the ETKF (which is $0.493$). As $n$ increases to $n=21$, the ETKF becomes to outperform the EnUKF. If $n$ increases further, the ETKF still outperforms the EnUKF, but the gap in their performances are narrowed. For example, the lowest relative rmse of the ETKF is $0.049$, which is obtained at $\delta=0.05$ with $n=61$, while the lowest relative rmse of the EnUKF is $0.075$, achieved at $\delta=0.05$ with $n=41$.

The above result reflects the difference between the EnUKF and the ETKF in their philosophies of statistics estimations. As shown in the Appendix of \cite{Luo-ensemble}, the unscented transform (UT) used in the EnUKF aims to produce the estimations which capture the first few terms in the Taylor expansions of the true mean of covariance, but which in general mismatch higher order terms. In contract, the Monte Carlo approximation adopted in the ETKF produces the estimations that converge asymptotically to the true statistics. However, given only a small ensemble size, the resulting large sampling errors may lead to large biases and spurious modes. For this reason, the relative superiority between the EnUKF and the ETKF may invert in the small and large sample scenarios.   

\section{Conclusion}

Through the supplementary experiment, we showed that, in the small sample scenario, the EnUKF can perform (slightly) better than the ETKF with the same ensemble size, while as the ensemble size increases, the ETKF outperforms the EnUKF instead. Our explanation of this phenomenon is the following: The estimations of the ETKF are based on the Monte-Carlo approximations, which converge asymptotically to the true values as the ensemble size increases. However, in the small sample scenario, the Monte-Carlo approximations may exhibit large biases and spurious modes due to the relatively large sampling errors, which are partially avoided in the estimations based on the UT. However, one possible problem of the UT is that its estimations may not converge to the true statistics as the ensemble size increases, since its ensemble members are deterministically chosen, rather than randomly drawn as the samples in the Monte-Carlo approximations. 


\section{Appendix: Implementation of the covariance filtering technique}

A comparison of the results presented in Figs.~\ref{fig:EnUKFL96_noFiltering_varInflationEnSize} and \ref{fig:ETKFL96_noFiltering_varInflationEnSize} (when $n=11$) in this reply and those in Figs.~2 and 3 of \cite{Luo-ensemble} confirms the benefit of covariance filtering. By conducting covariance filtering, one increases the rank of a sample covariance matrix. So in effect, it increases the effective ensemble size. Other benefits (and shortcomings) of covariance filtering are also discussed in, for example, \cite{Hamill2009,VanLeeuwen2009}.

Covariance filtering essentially involves computing the Schur product between a sample covariance and a taper matrix. Mathematically, the Schur product, $\mathbf{C} \equiv \mathbf{A} \circ \mathbf{B}$, of two matrices $\mathbf{A}$ and $\mathbf{B}$ with the same dimensions, is defined as the matrix with the same dimension as $\mathbf{A}$ and $\mathbf{B}$, whose components $C_{ij} = A_{ij} B_{ij}$, where $A_{ij}$, $B_{ij}$ and $C_{ij}$ are the components on the $i$th row and the $j$th column of the matrices $\mathbf{A}$, $\mathbf{B}$ and $\mathbf{C}$, respectively. In our discussion, we suppose that $\mathbf{A}$ is the covariance matrix, and $\mathbf{B}$ is the taper matrix used to reduce the spuriously large correlations in $\mathbf{A}$.

\begin{figure*}[!t]
\centering
\includegraphics[width=\textwidth]{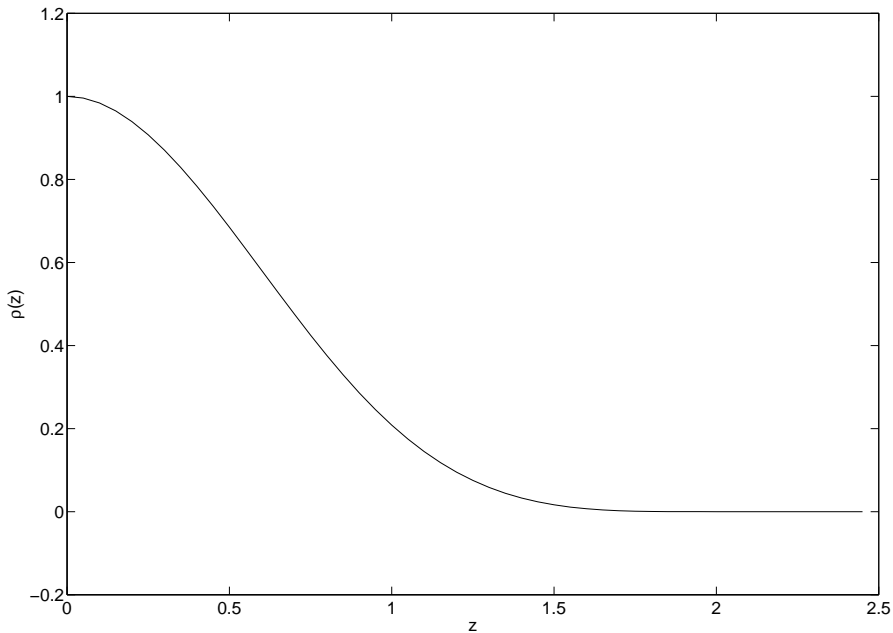}
\caption{ \label{fig:ch2_correlation_function} Shape of the function $\rho$ in Eq.~(\ref{ch2_cov_filtering_correlation_func}).}
\end{figure*}

The construction of $\mathbf{B}$ can be done in the following way: 
\begin{equation} \label{ch2:eq_taper matrix}
B_{ij} = \rho (d_{ij}),
\end{equation}
where $d_{ij}$ is a metric measuring the difference between the locations $i$ and $j$, and $\rho$ is a correlation function. Several examples of correlation functions were discussed in \cite{Gaspari1999}. In our experiments, we followed \cite{Houtekamer-sequential} and chose function $\rho$ in the following form:
\begin{equation} \label{ch2_cov_filtering_correlation_func}
\rho \left( z \right) = \begin{cases}
 -\dfrac{1}{4} z^5 + \dfrac{1}{2} z^3 + \dfrac{5}{8} z^3 - \dfrac{5}{3} z^2 +1 \, , & \text{if}~ 0 \le z \le 1 \, ; \\
\dfrac{1}{12} z^5 - \dfrac{1}{2} z^4 + \dfrac{5}{8} z^3 + \dfrac{5}{3} z^2 - 5 z + 4 - \dfrac{2}{3} z^{-1} \, , & \text{if}~ 1 < z \le 2 \, ;\\
 0 \, , & \text{if}~ z >2 \, . \\ 
\end{cases} 
\end{equation}  
For illustration, the shape of the function $\rho$ is plotted in Fig.~(\ref{fig:ch2_correlation_function}). As one can see there, the function $\rho$ has a ``cut-off'' effect at $z=2$, in the sense that the values of the function are set to zero for all $z>2$.    

For data assimilation in real world, $d_{ij}$ is normally a function of the distance between the locations $i$ and $j$ in the three dimensional physical world. For example, see \cite{Constantinescu2007}. However, in mathematical analysis, this choice might not always be available. For example, the system states of a mathematical model may not have any physical meaning, so that we cannot observe them in the physical world. On the other hand, for a mathematical model, the (physical) distance between  locations $i$ and $j$ may also not be well defined as in the physical world. 

For the above reasons, in our experiments we chose to calculate $d_{ij}$ based on the covariance matrix $\mathbf{A}$ itself. Suppose that $\mathbf{A}$ is a matrix with $m$ rows (and the number of its rows is not less than the number of its columns), so that 
\begin{equation} \label{ch2:cov_before_tapering}
\mathbf{A} = \left [ \mathbf{r}_1^T, \dotsb, \mathbf{r}_m^T \right ]^T,
\end{equation}
where $\mathbf{r}_i$ is the $i$th row of $\mathbf{A}$. We define
\begin{equation} \label{ch2:distance_in_column_vectors}
d_{ij}=\lVert \mathbf{r}_{i}^T-\mathbf{r}_{j}^T \rVert_{2} / l_c \, ,
\end{equation}  
where $l_c$ is a length scale that is introduced to influence where the ``cut-off'' effect of the function $\rho$ takes place\footnote{Note that we have assumed that the number of the rows of a matrix (not necessarily square) is not less than the number of its columns. If this is not the case, then it is suggested to choose the column vectors to calculate the distances $d_{ij}$ in Eq.~(\ref{ch2:distance_in_column_vectors}) instead. In this way, covariance filtering can be applied to non-square matrices like the cross covariance (when the dimension of the state space is not equal to that of the observation space).}. But note that here, $l_c$ does not have any physical meaning (i.e., it does not correspond to any quantity in the physical world). Instead, it is interpreted as a threshold of the statistical metric $d_{ij}$, which measures the (statistical) difference between the $i$th and $j$th elements of a random vector $\mathbf{x}$. In \cite[\S~3.3.3.2]{LuoXDThesis}, one of us showed through a numerical example that covariance filtering conducted in this way can achieve the same effect as those reported in the literature by using the physical distances between different locations to construct the taper matrix. 

In \cite{Luo-ensemble}, covariance filtering was conducted on the background covariance, the cross covariance, and the projection covariance (cf. Eq.(22) of \cite{Luo-ensemble}), since all of them are involved in updating the background to the analysis (cf. cf. Eq.(23) of \cite{Luo-ensemble}). No covariance filtering was conducted on the quantities other than the above three covariance matrices.

\bibliographystyle{elsart-num-sort}
\bibliography{references}

\begin{thebibliography}{10}
\expandafter\ifx\csname url\endcsname\relax
  \def\url#1{\texttt{#1}}\fi
\expandafter\ifx\csname urlprefix\endcsname\relax\def\urlprefix{URL }\fi

\bibitem{Constantinescu2007}
E.~M. Constantinescu, A.~Sandu, T.~Chai, G.~R. Carmichael, Ensemble-based
  chemical data assimilation. ii: {C}ovariance localization, Quart. J. Roy.
  Meteor. Soc. 133 (2007) 1245--1256.

\bibitem{Gaspari1999}
G.~Gaspari, S.~E. Cohn, Construction of correlation functions in two and three
  dimensions, Quart. J. Roy. Meteor. Soc. 125 (1999) 723 -- 757.

\bibitem{Hamill2009}
T.~M. Hamill, J.~S. Whitaker, J.~L. Anderson, C.~Snyder, Comments on
  ``{S}igma-point {K}alman filter data assimilation methods for strongly
  nonlinear systems'', Journal of the Atmospheric Sciences 66 (2009)
  3498--3500.

\bibitem{Houtekamer-sequential}
P.~L. Houtekamer, H.~L. Mitchell, A sequential ensemble filter for atmospheric
  data assimilation, Mon. Wea. Rev. 129 (2001) 123--137.

\bibitem{LuoXDThesis}
X.~Luo, Recursive {B}ayesian filters for data assimilation, Ph.D. thesis,
  University of Oxford (2010).
\newline\urlprefix\url{http://arxiv.org/abs/0911.5630}

\bibitem{Luo-ensemble}
X.~Luo, I.~M. Moroz, Ensemble {K}alman filter with the unscented transform,
  Physica D 238 (2009) 549--562.

\bibitem{Ott-local}
E.~Ott, B.~R. Hunt, I.~Szunyogh, A.~V. Zimin, E.~J. Kostelich, E.~J. Corazza,
  E.~Kalnay, D.~J. Patil, J.~A. Yorke, A local ensemble {K}alman filter for
  atmospheric data assimilation, Tellus 56A (2004) 415--428.

\bibitem{Sakov2009}
P.~Sakov, Comment on ``{E}nsemble {K}alman filter with the unscented
  transform'', Physica D 238 (2009) 2227--2228.

\bibitem{Sakov2008}
P.~Sakov, P.~R. Oke, Implications of the form of the ensemble transformation in
  the ensemble square root filters, Mon. Wea. Rev. 136 (2008) 1042--1053.

\bibitem{VanLeeuwen2009}
P.~J. Van~Leeuwen, Particle filtering in geophysical systems, Mon. Wea. Rev.
  137 (2009) 4089--4114.

\bibitem{Wang-which}
X.~Wang, C.~H. Bishop, S.~J. Julier, Which is better, an ensemble of
  positive-negative pairs or a centered simplex ensemble, Mon. Wea. Rev. 132
  (2004) 1590--1605.

\bibitem{Whitaker-ensemble}
J.~S. Whitaker, T.~M. Hamill, Ensemble data assimilation without perturbed
  observations, Mon. Wea. Rev. 130 (2002) 1913--1924.

\end{thebibliography}
\end{document}